# Terahertz spin currents in nanoscale spatial resolution

Jiahua Cai[1, ‡], Mingcong Dai[1, ‡], Sai Chen[1, ‡, *], Peng Chen[2], Jiaqi Wang[1], Hongting Xiong[1] Zejun Ren[1], Shaojie Liu[3], Zhongkai Liu[4, 5], Caihua Wan[2] and Xiaojun Wu[1, 6, 7, *]

[1] School of Electronic and Information Engineering, Beihang University, Beijing, 100191, PR China.

[2] Institute of Physics, Chinese Academy of Science, Beijing, 100190, PR China.

[3] School of Cyber Science and Technology, Beihang University, Beijing, 100191, PR China.

[4] School of Physical Science and Technology, ShanghaiTech University, Shanghai, 201210, PR China.

[5] ShanghaiTech Laboratory for Topological Physics, ShanghaiTech University, Shanghai, 201210, PR China.

[6] Zhangjiang Laboratory, 100 Haike Road, Shanghai, 201210, PR China.

[7] Wuhan National Laboratory for Optoelectronics, Huazhong University of Science and Technology, Wuhan 430074, PR China.

*Corresponding authors. E-mail: Sai Chen, saichen@buaa.edu.cn. Xiaojun Wu, xiaojunwu@buaa.edu.cn

‡These authors contributed equally to this study.

# Highlights

- Spintronic THz emission nanoscopy (STEN) permits the nanoscale injection and detection of ultrafast THz spin current.
- STEN is demonstrated to be an effective technique for inspecting the integrity of spintronic heterostructure films.
- Due to the intense localised field enhancement afforded by the nano-tip, the surface of spin heterostructures can be fabricated and characterised at higher pump powers.

**TOC**

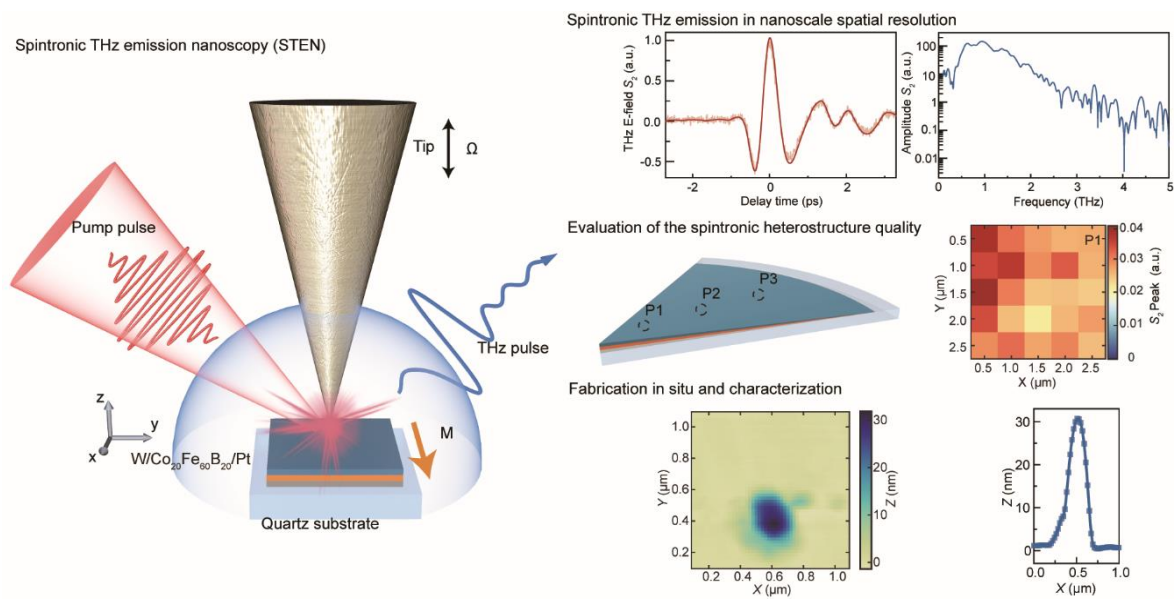

## Abstract

The ability to generate, detect, and control coherent terahertz (THz) spin currents with femtosecond temporal and nanoscale spatial resolution has significant ramifications. The diffraction limit of concentrated THz radiation, which has a wavelength range of 5 μm-1.5 mm, has impeded the accumulation of nanodomain data of magnetic structures and spintronic dynamics despite its potential benefits. Contemporary spintronic optoelectronic apparatuses with dimensions 100 nm presented a challenge for researchers due to this restriction. In this study, we demonstrate the use of spintronic THz emission nanoscopy (STEN), which allows for the efficient injection and precise coherent detection of ultrafast THz spin currents at the nanoscale. Furthermore, STEN is an effective method that does not require invasion for characterising and etching nanoscale spintronic heterostructures. The cohesive integration of nanophotonics, nanospintronics, and THz-nano technology into a single platform is poised to accelerate the development of high-frequency spintronic optoelectronic nanodevices and their revolutionary technical applications.

# 1 Introduction

In the post-Moore's-Law era, spintronic is anticipated to satisfy the stringent criteria for storage density and fast read/write rates. This is the primary reason why spintronic has recently garnered so much attention[1,2]. The study of spintronic material[3,4] processes, structures, devices, and other fields has advanced significantly, and it is now progressively moving from the laboratory research stage to commercial applications. For example, two-dimensional magnetic materials have been fabricated into heterostructures since 2016[5–9]. It can be considered to have been made vertically thinner, resulting in thinner devices that meet the requirements. Some studies, on the other hand, focus on terahertz (THz) frequency range spin currents to attain switch speeds close to THz frequencies[10–16]. Thus, by employing a femtosecond laser-induced THz emission approach[10], THz spin currents were produced and contactless observed. This method has been not only a potent instrument for spintronic material research[12,13,17], but also a possible alternative to Lithium Niobate[18] as an ultrafast intense source[19]. However, the aforementioned studies do not address the crucial issue of spatial resolution at THz frequencies, which is growing in importance as storage capacities expand. If a system's controlling dimensions approach the nanoscale, the viability of spin currents and THz frequency range operation must be investigated.

Scattering scanning near-field optical microscopy (s-SNOM) has emerged as a potent method for characterization of nanoscale materials and devices[20–27]. In recent years, THz-coupled s-SNOM (THz s-SNOM) has emerged as a useful instrument for studying the response of the THz frequency band at the nanoscale with sub-picosecond temporal resolution[28–35]. Its spatial resolution can achieve a multiplication of $\lambda/10000$, where $\lambda$ is the wavelength, which is more crucial than in optical SNOM and offers better dimensional complementarity[36–40]. Additionally, the introduction of the transient optical pump-THz probe[41] (OPTP) in SNOM allows for the investigation of ultrafast nanoscale dynamics, especially in two-dimensional materials. Moreover, a study conducted by Daniel Mittleman and colleagues in 2017 utilized nanoscale ultrafast laser-induced THz emission nanoscopy (LTEN) to observe THz emission in InAs[42]. In 2021, T. L. Cocker and R. Huber et al. conducted a study that showcased the potential of THz emission as a means to examine the tunneling current in $WSe_2/WS_2$ heterostructures[43]. Their findings indicated that the signal-to-noise ratio of the THz emission spectrum had significantly improved. However, this methodology has not yet been observed to be utilized in the generation, detection, and manipulation of femtosecond spin currents.

In this study, nanometer-resolution THz spin current pulse was initially observed from the $W/Co_{20}Fe_{60}B_{20}/Pt$ heterostructure[19,44,45], utilizing spintronic THz emission nanoscopy (STEN) based on LTEN. We have demonstrated that the THz radiation emitted by the nanoscale spin current is still following the Inverse Spin Hall Effect (ISHE)[10] and has a high radiation efficiency, with the fifth harmonic order of emission signals being visible. Notably, we can observe well-resolved THz emission signals even with a modest pump power of 1 mW or from two-layer $Co_{20}Fe_{60}B_{20}/Pt$ films which promise this method can be broadly applied in many cases. Meanwhile, STEN is proved to be a powerful approach for the quality inspection of spintronic heterostructure films. Furthermore, owing to the substantial enhancement of the localized field provided by the nano-tip, we were capable of fabricating the sample surface in situ at elevated pump powers.

## 2 Experimental Section

### 2.1 STEN (Spintronic THz emission nanoscopy)

The experimental setup utilized in this study is founded upon a commercially accessible s-SNOM system that employs an AFM, manufactured by Neaspec GmbH in Germany. The AFM tip is subjected to illumination through a fiber-coupled THz time-domain spectroscopy system (Terasmart[46], Menlo GmbH, Germany) and an external femtosecond laser light. The laser light employed exhibits a central wavelength of 780 nm, a repetition frequency of 100 MHz, a pulse duration of 100 fs, and a maximum power output of 75 mW. When exclusively conducting THz-TDS operations, the setup operates in the capacity of a THz scattering-type scanning near-field optical microscope (s-SNOM). The OPTP mode configuration functions by employing both THz-TDS and external laser light. In the event that the THz-TDS's THz emitter is non-functional, the configuration will function in LTEN mode. The STEN functions in the LTEN operational mode. Additional information can be found in the Supplementary Information section.

**2.2 Preparation of Ferromagnetic-heavy metal heterostructure sample**

Magnetron sputtering was used to grow the W/$Co_{20}Fe_{60}B_{20}$/Pt heterostructures[15] on a double-polished quartz substrate, which is ideal for THz light transmission through magnetic thin films. The growth took place at room temperature, with a chamber base pressure below $2\times10^{-8}$ Torr. An in-plane external magnetic field of 180 Oe was applied to induce exchange bias or coupling. The Pt/$Co_{20}Fe_{60}B_{20}$/W heterostructures stacks underwent the same process.

## 3 Results and Discussion

**3.1 Spintronic THz emission in nanoscale spatial resolution**

To assess whether THz spin currents could be excited with nanoscale spatial resolution, we performed experiments using LTEN (Ultrafast THz s-SNOM[47], Neaspec GmbH, Germany) to measure the THz emission from the W/$Co_{20}Fe_{60}B_{20}$/Pt heterostructure, which has been demonstrated to exhibit the best emission in prior studies. A schematic diagram of the experiment is shown in Fig. 1a, and b. The AFM's tip has a diameter of approximately 50 nm and oscillates at around 48 kHz. The heterostructure was pumped with a femtosecond laser having a central wavelength of 780 nm, a duration of 100 fs, a repetition frequency of 100 MHz, and a power of 14.5 mW. The THz wave was detected using a photoconductive antenna that was excited by the same femtosecond laser at a power of 20.7 mW, and an integration time of 500 ms was set. The experiment was conducted in ambient air, and the resulting THz temporal waveforms obtained from the STEN are depicted in Fig. 1c. The Figure presents the 2nd harmonic demodulation signal of the tip oscillation frequency (solid line) fitted from actual experimental data (shallow line), which closely resembles the THz waveform measured using the THz emission measuring system in an open space. The frequency spectrum corresponding to the waveform is shown in Fig. 1d. The photoconductive antenna limited the bandwidth, and the water absorption peak is visible, as the STEN is exposed to the open environment. Furthermore, the third, fourth, and fifth harmonic signals were detected and are presented in Fig. 1e, indicating a sufficient signal-to-noise ratio. Based on these experimental results, we successfully identified the THz waves that could be excited from the $W/Co_{20}Fe_{60}B_{20}/Pt$ heterostructure with a spatial resolution of at least 50 nm.

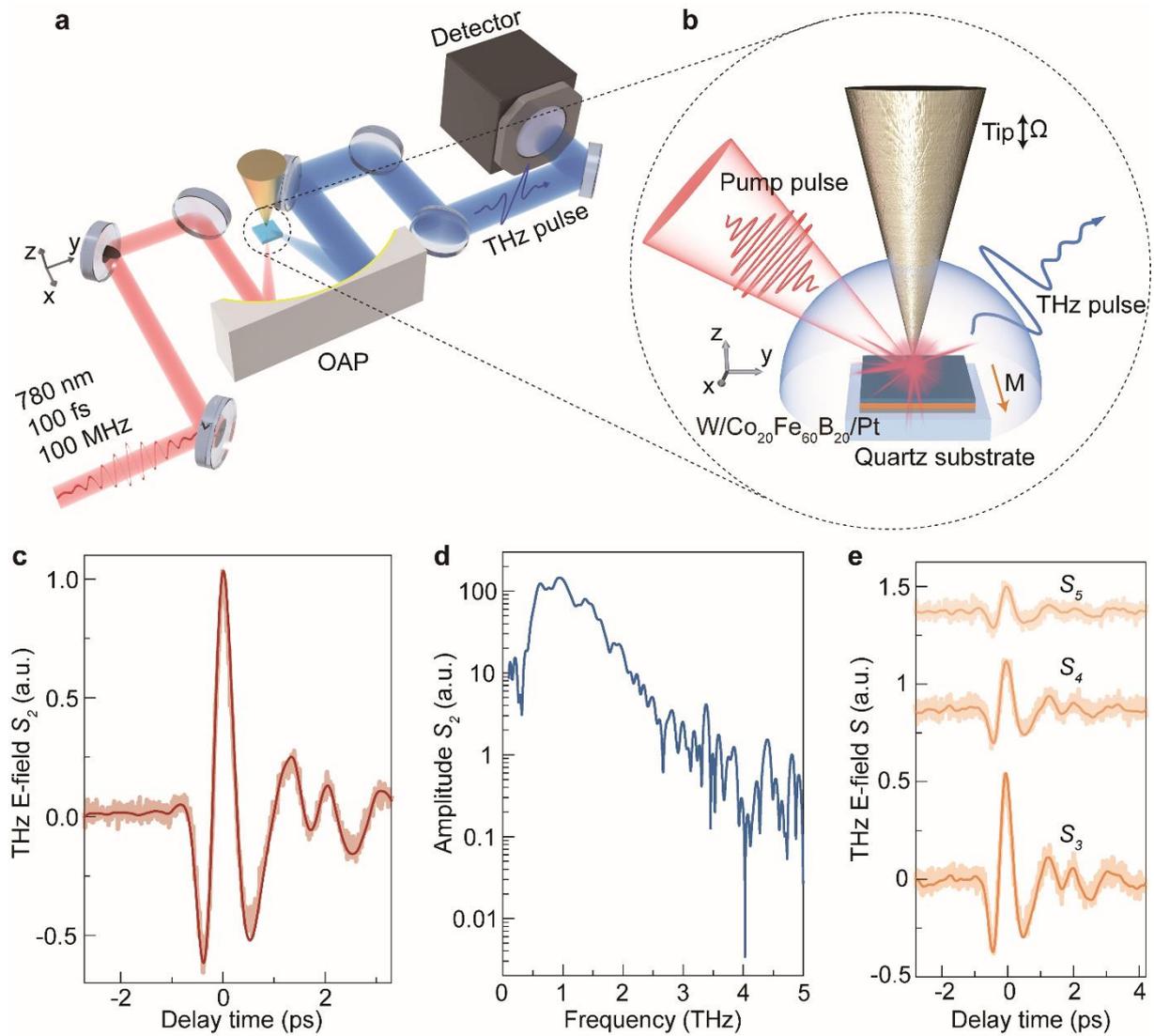

**Fig. 1** Injection and detection of nanoscale THz spin currents at ultra-low pump power, and nanoscale THz emission of femtosecond laser pumped ferromagnetic/non-magnetic metal heterostructures. **a** Schematic of the STEN. An optical pulse (pulse duration <100 fs, central wavelength 780 nm, repetition rate 100 MHz), coupled to the nanoscale tip, is used for ultrafast pumping of the W/Co$_{20}$Fe$_{60}$B$_{20}$/Pt heterostructure and generating the THz pulses which can be detected by a photoconductive antenna. OAP: off axis parabolic mirror; Ω: tip oscillation frequency. **b** Schematic of the optical pulse pump and the THz emission. **c** Pump-induced (pump power 14.5 mW) 2$^{nd}$ THz harmonic demodulation waveforms recorded on the W/Co$_{20}$Fe$_{60}$B$_{20}$/Pt heterostructure as a function of the THz delay time. The solid red line represents the fitted result and the shallow line is the original result. **d** Relative laser-induced

spectral amplitude for the 2$^{nd}$ harmonic signal. **e** Different THz emission signals of 3$^{rd}$, 4$^{th}$, and 5$^{th}$ harmonic demodulation

To determine whether the radiated THz waves are coming from the nanoscale THz spin currents, we further conduct several experiments as shown in Fig. 2. In the conversion from THz spin currents to THz radiation, the coherent THz electromagnetic waves are radiated via the ISHE, and the THz emission is related to the spin Hall angle $\gamma$ and magnetization $\vec{M}$ ($\vec{E}_{THz} \propto \boldsymbol{\gamma} \cdot \vec{j}_S \times \frac{\vec{M}}{|\vec{M}_S|}$). Hence, to determine whether the radiated THz waves originate from the THz spin currents, one possible approach is to modify variables $\boldsymbol{\gamma}$, $\vec{j}_S$ and $\vec{M}$, subsequently, observe changes in the THz electric field. Determining the origin of the radiated THz wave in free space requires an assessment based on sample flipping; however, this method is not suitable for STEN. Fig. 2b illustrates that, to address this issue, we measured the W/Co$_{20}$Fe$_{60}$B$_{20}$/Pt heterostructure deposited with both sequence and reverse order[48] in STEN, and the waveform in the time domain is flipped. It is direct evidence that the THz radiated waves originate from the THz spin currents.

Furthermore, we examined the two-layer heterostructure Co$_{20}$Fe$_{60}$B$_{20}$/Pt. In contrast to the scenario in free space, the THz electric field difference between the two-layer and three-layer heterostructures is not noticeable. This could be due to the localized surface plasmon resonance induced by the tip, which causes the spintronic heterostructure to absorb mostly femtosecond lights and generate THz spin currents with comparable amplitude. In addition, the signals were analyzed by rotating the azimuthal angle of the sample while maintaining its saturation magnetization. Fig. 2c demonstrates that despite only collecting z-polarized THz waves in STEN, the behavior of THz emission in the spintronic heterostructures resembles that in free space. The complete set of experimental data can be located in the Supplementary Information. These results present compelling evidence that the THz emission originating from the spintronic heterostructure can be interpreted as an electric dipole that emits THz waves uniformly in all directions while exhibiting distinct polarization characteristics. In free space, the THz intensity of the spintronic heterostructure does usually not reach saturation with the increase of the pump power. However, concerning Fig. 2d, it can be observed that the THz electric fields in STEN manifest a linear correlation with the pump power up to 10 mW, beyond which it tends towards saturation, as evidenced by the 12 mW pump power level. Figure 2e depicts the fitted THz emission at a pump power of 1 mW. Even at 0.5mW pump power, we can measure the THz emission signal described in the Supplementary Information section.

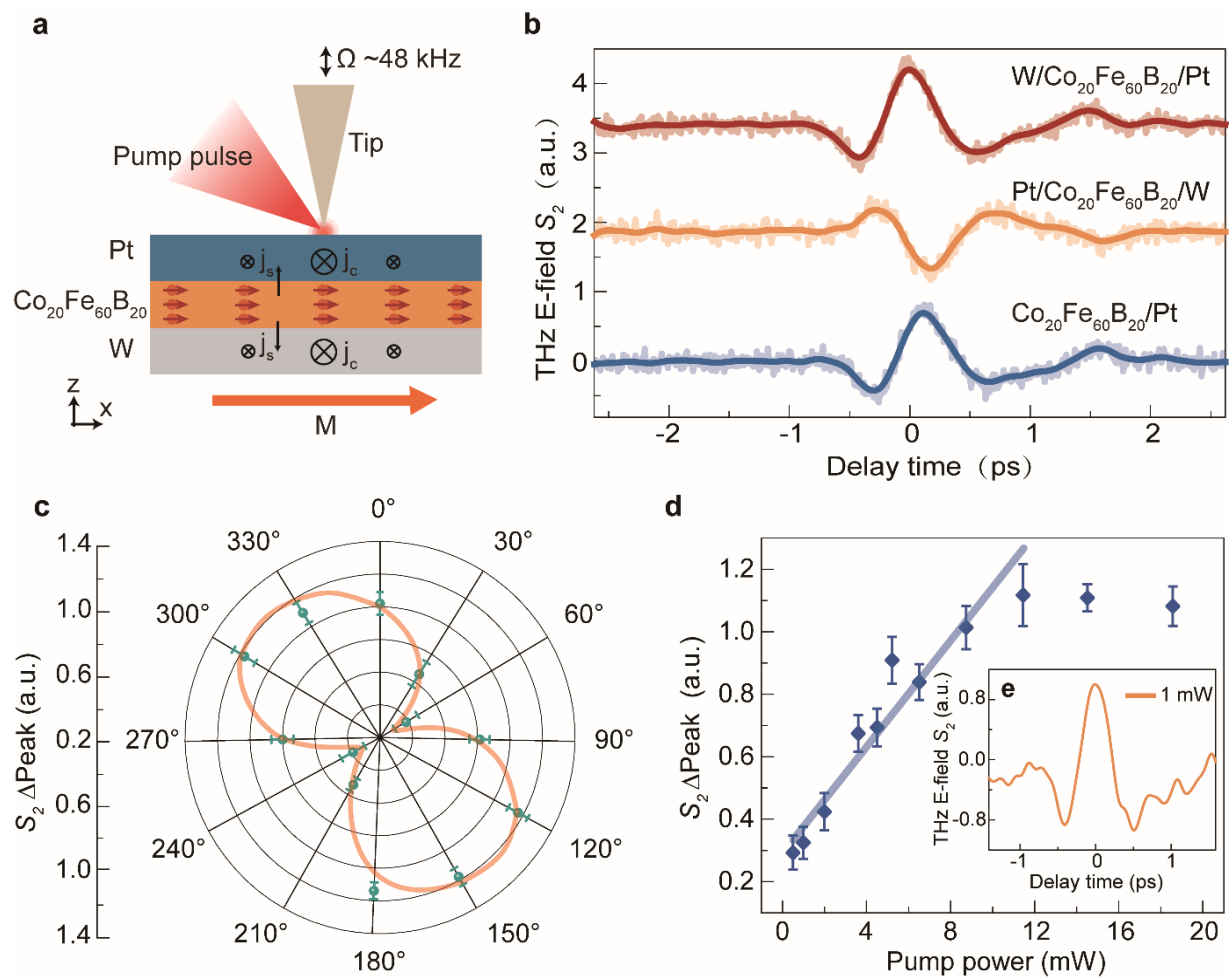

**Fig. 2** ISHE occurring in the nanoscale W/$Co_{20}Fe_{60}B_{20}$/Pt heterostructures. **a** Schematic of the ISHE. **b** The 2$^{nd}$ harmonic THz emission time domain waveforms (the solid line) and the fitted result (the shallow line) from the W/$Co_{20}Fe_{60}B_{20}$/Pt heterostructures deposited with sequence

and reverse order, and the result from the two-layer heterostructure $Co_{20}Fe_{60}B_{20}$/Pt. **c** Peak-to-peak values and fitting results of the THz emission signals obtained at different sample azimuths. **d** The peak-to-peak values of THz emission signals at different pump powers. The solid line shows a linear fit to the data. **e** The fitted transient at the pump power of 1 mW

### 3.2 Spatial resolution measurement

To evaluate the efficacy of STEN in assessing the quality of spintronic heterostructures, a specific fragment located at the periphery of the sample is chosen for measurement. Fig. 3a displays that in the STEN, measurement was performed on three different positions (P1, P2, and P3). The 2$^{nd}$ harmonic demodulation scattered THz electric field is illustrated in Fig. 3b. The corresponding topography and near-field THz emission signal's peak maps are shown in Fig. 3c-e, and f-h, respectively. As is typical for magnetron sputtering, the thicknesses of P1, P2, and P3 decrease as the position moves closer to the edge. Regarding topographical maps, a clear distinction can be made between three different positions based on the average thickness difference of 16 nm. Likewise, the mapping of THz emission at near-field range exhibits similar patterns, wherein the mean variation of the emitted THz field across the three locations enables straightforward differentiation of positioning. The Figures exhibit a resolution of 500 nm, which is attributed to the constraints of signal acquisition. However, it is possible to achieve a resolution as minimal as 50 nm. It should be observed that the thickness of the sputtering film in the W/$Co_{20}Fe_{60}B_{20}$/Pt heterostructure is 6 nm, a value significantly lower than the measurement obtained from AFM topography. The lack of uniformity evident in topography images may be ascribed to the inclined positioning of the sample and the roughness of the substrate surface. This suggests that relying exclusively on AFM topography for evaluating the excellence of sputtering heterostructure may present difficulties. On the other hand, the STEN metric provides a direct reflection of the spintronic performance, which accurately indicates the quality of spintronic heterostructures. Consequently, STENs can function as effective and non-contact tools for characterizing spintronic devices with nano-resolution.

Furthermore, we performed simultaneous optical pump - THz probe (OPTP) experiments, as shown in Fig. 3i. The graph illustrates how the peak-to-peak value of the scattering signal $S_2$ of the THz-TDS changes as the delay time of the optical pump is varied. The Supplementary Information section contains the entire scattering signal of the THz-TDS. Fig. 3j presents the $S_2$ data for the emission time-domain spectroscopy under the same conditions, and Fig. 3k shows the subtraction of the two signals, providing a nearly straight line that indicates that the THz emission is primarily responsible for the variations. This result demonstrates the inherent advantage of STEN over OPTP for spintronic THz emission. Typically, an increase in a film conductivity, such as a semiconductor, would lead to significant changes in its optical properties that would affect OPTP signals, but light-induced currents ($\vec{j}_C$) are unable to influence OPTP signals. In those cases, OPTP signals usually reflect conductivity variations. However, as the conductivity is already high, light-induced conductivity variations have an insignificant impact on the optical performance of the s-SNOM. Thus, most OPTP signals originate from $\vec{j}_C$, predominantly derived from spin conversion currents ($\vec{j}_{SCC}$).

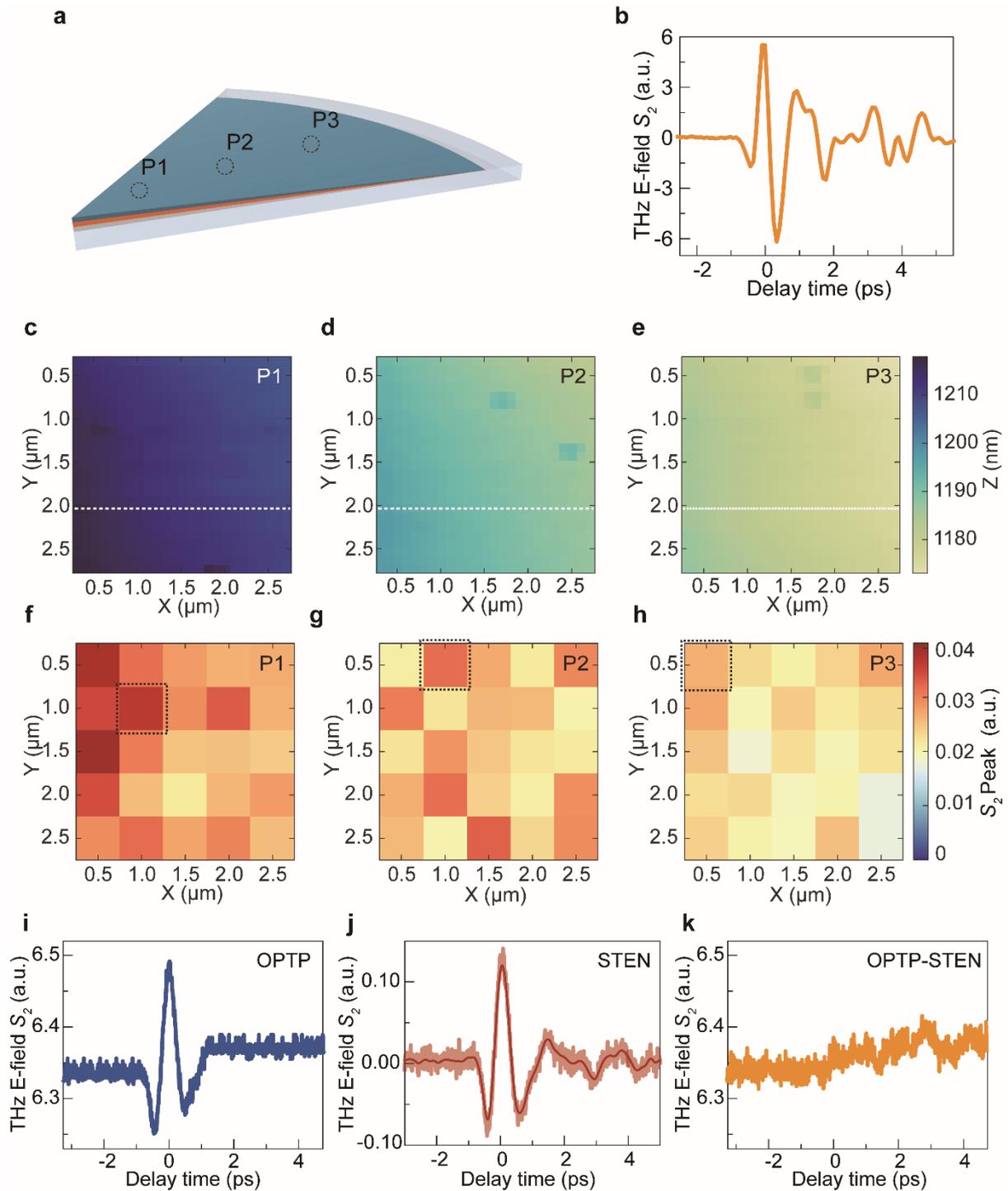

**Fig. 3** THz emission-based evaluation of the spintronic heterostructure quality. **a** The schematic diagram of sample layer thickness, where P1, P2, and P3 are three different positions that gradually approach the edge and decrease in layer thickness. **b** The 2$^{nd}$ harmonic demodulation scattered the electric field signal from the sample surface. The 2.5×2.5 μm range of the AFM topography at P1 (**c**), P2 (**d**), and P3 (**e**), and the THz emission peak plots at positions P1 (**f**),

P2 (**g**), and P3 (**h**). **i** The THz emission signal extracted from OPTP. **j** The pump-induced THz emission waveform. **k** The signal waveform after subtraction between the OPTP signal and the emission signal

**3.3 Fabrication in situ and characterization**

As depicted in Fig. 2d, the THz emission signal attains saturation when the pump power reaches 12 mW due to the highly enhanced field from the tip. However, there is a slight reduction in the emission signal when the power is further increased to 20 mW, indicating possible degradation of the film due to the intensely concentrated light field. This observation piqued our interest, leading to the measurement of the pump power dependence of the STEN signal ($S_2$) using tips of varying radii (57 nm, 67 nm, and 207 nm) as shown in Fig. 4a. The peak-peak value increases with an increase in the tip radius and begins to drop at 6.5 mW, 8.7 mW, and 11.5 mW, respectively. We note that the tip radius affects the light field enhancement[31], which is different for the pump light and THz waves because of their distinct wavelengths. In this instance, the THz wave enhancement is more significant, leading to a much higher emission signal from the 207-nm-radius tip. However, there is little difference in the drop point of the pump power. To investigate what occurred at the point where the emission signal dropped, we conducted AFM topography mapping over an area of 3×3 μm around that region, as shown in Fig. 4c-e. The results reveal that the pumping light induces visible protrusions. Supplementary Information provides the corresponding s-SNOM mapping. Comparing the spatial profiles of the protrusions from the two methods yielded strikingly similar results. In Fig. 4b, the cross-section profile of three protrusions is displayed, indicating that a larger radius tip results in a greater bulge height ranging from 1.6 nm to 3.5 nm. Additionally, the spatial profile is increased from 380 nm to 530 nm. In Fig. 4f, and g, we utilized a tip with a radius of ~180 nm and set the pump power to 20 mW, presenting the AFM topography and the THz scattering image, respectively. The spatial profile of the bulge remains at 500 nm. The cross-section profile at Y=0.4 μm is depicted in Fig. 4h, with height variation reaching 30 nm. Our study results reveal that STEN provides the potential for on-site production of spintronic films with nanometer accuracy, representing a vital aspect in the advancement of precise manufacturing techniques for magnetic computing and storage devices.

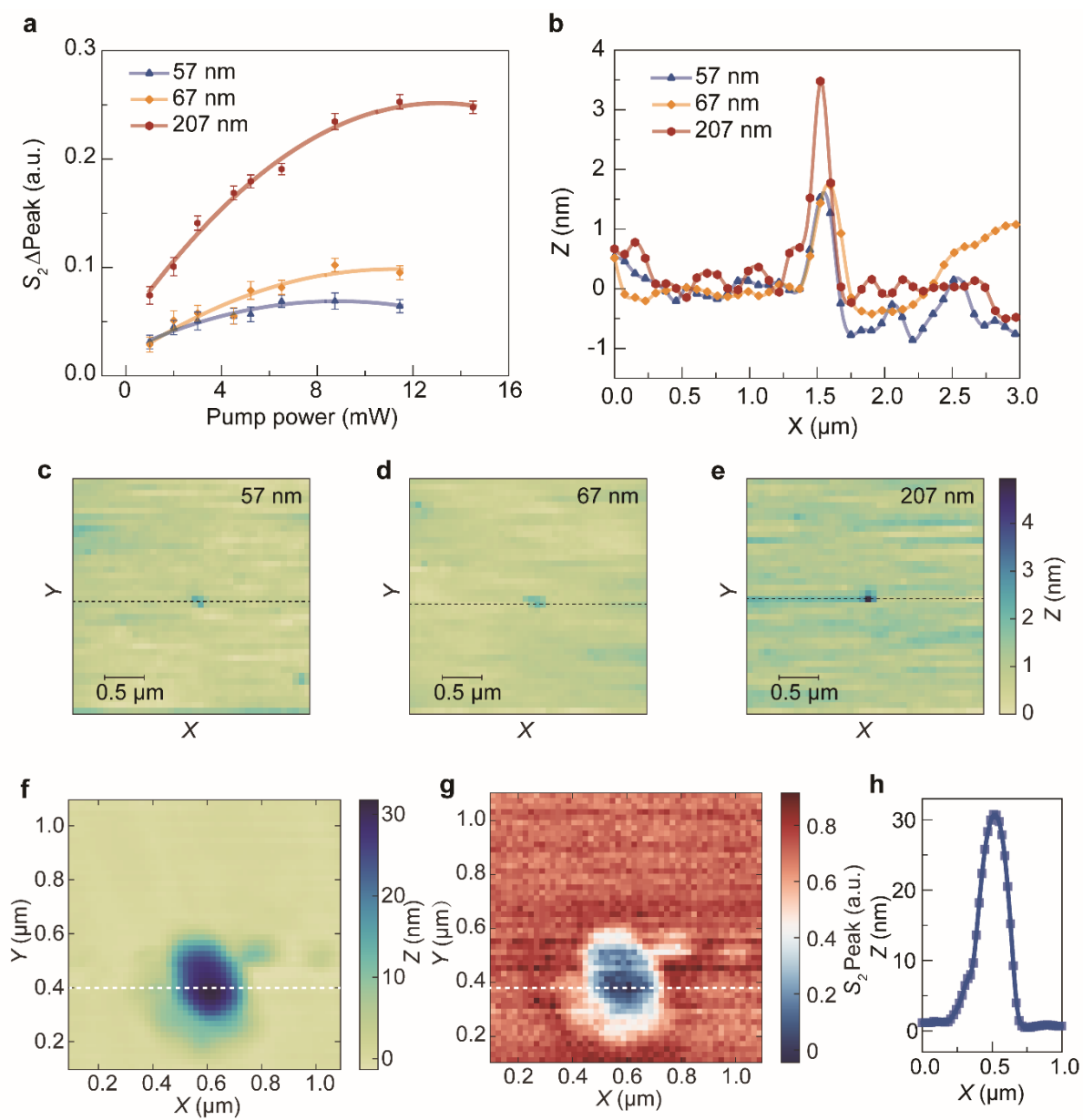

**Fig. 4** The sample surface was etched by the tip-guided pump laser. **a** Power dependence with tips radius of 57 nm, 67 nm, and 207 nm. **b** The height of protrusions on the surface of the sample, caused by different radius tips burning, varies with the X-axis corresponding to the white dashed line positions in **c-e**. AFM topography of the sample surface after the action of a tip with a radius of 57 nm (**c**), 67 nm (**d**), and 207 nm (**e**). AFM topography **f** and THz scattering images **g** of protrusions left by burning on the surface of the sample under the action with the tip. **h** The height variation curve of the sample surface along the white dashed line position in **f** and **g**

## 4  Conclusions

The present study reports the preliminary detection of THz spin current pulses with nanometer resolution from W/Co$_{20}$Fe$_{60}$B$_{20}$/Pt heterostructures, employing LTEN via an ultrafast THz s-SNOM. The results suggest that the production of THz radiation through nanoscale spin current aligns with the principles of ISHE and demonstrates a significant level of radiation efficiency. The presence of emission signals at the fifth harmonic order suggests a signal-to-noise ratio of a considerable magnitude. Using a comparatively low pump power of 1 mW, the tri-layer sample emits signals with a high resolution. Additionally, the two-layer Co$_{20}$Fe$_{60}$B$_{20}$/Pt heterostructures can emit a detectable emission signal. This discovery suggests that this technique has the potential for widespread applicability. In addition, due to the substantial enhancement of the localized field provided by the nano-tip, we are able to modify the surface of the sample in situ using high pump powers. The findings suggest that STEN could be used as a non-invasive technique for the characterization and fabrication of nanoscale spintronic heterostructures, a crucial factor in the development of high-density storage devices.

## Ethics Declarations
### Conflict of Interest
The authors declare no interest confict. They have no known competing financial interests or personal relationships that could have appeared to influence the work reported in this paper.

## Acknowledgments

Supported by the National Key R&D Program of China (2022YFA1604402), the National Natural Science Foundation of China (11827807, 92250307, 62005140), Technology Innovation Action Plan of the Science and Technology Commission of Shanghai Municipality with project number 20JC1416000, the Open Project Program of Wuhan National Laboratory for Optoelectronics NO.2022WNLOKF006.

## Supplementary Information

The overall system of the THz-TDS, OPTP, and LTEN, scattering THz signals of Au and W/$Co_{20}Fe_{60}B_{20}$/Pt heterostructure, the dependence of sample azimuth on THz emission signal, the THz emission signal at a pump power of 0.5mW, THz emission waveforms at various sample positions, and the effect of tip state on experimental results are discussed.

## Author Contributions

J. C., M. D., and S. C. contributed equally to this study. X. W. and S. C. conceived and coordinated the STEN project. X. W., S. C., J. C., and M. D. designed the experimental setup and performed the experiments. S. C., X.W., M.D., and J.C. wrote the manuscript and revised the input from all.

**Figure and table captions**

**Fig. 1** Injection and detection of nanoscale THz spin currents at ultra-low pump power, and nanoscale THz emission of femtosecond laser pumped ferromagnetic/non-magnetic metal heterostructures. **a** Schematic of the STEN. An optical pulse (pulse duration <100 fs, central wavelength 780 nm, repetition rate 100 MHz), coupled to the nanoscale tip, is used for ultrafast pumping of the W/$Co_{20}Fe_{60}B_{20}$/Pt heterostructure and generating the THz pulses which can be detected by a photoconductive antenna. OAP: off axis parabolic mirror; Ω: tip oscillation

frequency. **b** Schematic of the optical pulse pump and the THz emission. **c** Pump-induced (pump power 14.5 mW) $2^{nd}$ THz harmonic demodulation waveforms recorded on the W/$Co_{20}Fe_{60}B_{20}$/Pt heterostructure as a function of the THz delay time. The solid red line represents the fitted result and the shallow line is the original result. **d** Relative laser-induced spectral amplitude for the $2^{nd}$ harmonic signal. **e** Different THz emission signals of $3^{rd}$, $4^{th}$, and $5^{th}$ harmonic demodulation

**Fig. 2** ISHE occurring in the nanoscale W/$Co_{20}Fe_{60}B_{20}$/Pt heterostructures. **a** Schematic of the ISHE. **b** The $2^{nd}$ harmonic THz emission time domain waveforms (the solid line) and the fitted result (the shallow line) from the W/$Co_{20}Fe_{60}B_{20}$/Pt heterostructures deposited with sequence and reverse order, and the result from the two-layer heterostructure $Co_{20}Fe_{60}B_{20}$/Pt. **c** Peak-to-peak values and fitting results of the THz emission signals obtained at different sample azimuths. **d** The peak-to-peak values of THz emission signals at different pump powers. The solid line shows a linear fit to the data. **e** The fitted transient at the pump power of 1 mW

**Fig. 3** THz emission-based evaluation of the spintronic heterostructure quality. **a** The schematic diagram of sample layer thickness, where P1, P2, and P3 are three different positions that gradually approach the edge and decrease in layer thickness. **b** The $2^{nd}$ harmonic demodulation scattered the electric field signal from the sample surface. The 2.5×2.5 μm range of the AFM topography at P1 (**c**), P2 (**d**), and P3 (**e**), and the THz emission peak plots at positions P1 (**f**), P2 (**g**), and P3 (**h**). **i** The THz emission signal extracted from OPTP. **j** The pump-induced THz emission waveform. **k** The signal waveform after subtraction between the OPTP signal and the emission signal

**Fig. 4** The sample surface was etched by the tip-guided pump laser. **a** Power dependence with tips radius of 57 nm, 67 nm, and 207 nm. **b** The height of protrusions on the surface of the sample, caused by different radius tips burning, varies with the X-axis corresponding to the white dashed line positions in **c-e**. AFM topography of the sample surface after the action of a tip with a radius of 57 nm (**c**), 67 nm (**d**), and 207 nm (**e**). AFM topography **f** and THz scattering images **g** of protrusions left by burning on the surface of the sample under the action with the tip. **h** The height variation curve of the sample surface along the white dashed line position in **f** and **g**